\documentclass[twocolumn]{aastex631}
\usepackage{booktabs}
\usepackage[flushleft]{threeparttable}
\usepackage{tabularx}
\usepackage{comment}

\newcommand{\rcn}{$\rm ^{12}CN/^{13}CN$}

\newcommand{\icn}{$\rm ^{12}CN$}
\newcommand{\jcn}{$\rm ^{13}CN$}

\submitjournal{ApJ}

\begin{document}

\title{The First Spatially-resolved Detection of \jcn\ in a Protoplanetary Disk and Evidence for Complex Carbon Isotope Fractionation}

\author[0000-0001-8002-8473]{Tomohiro C. Yoshida}
\email{tomohiroyoshida.astro@gmail.com}
\affiliation{National Astronomical Observatory of Japan, 2-21-1 Osawa, Mitaka, Tokyo 181-8588, Japan}
\affiliation{Department of Astronomical Science, The Graduate University for Advanced Studies, SOKENDAI, 2-21-1 Osawa, Mitaka, Tokyo 181-8588, Japan}

\author[0000-0002-7058-7682]{Hideko Nomura}
\affiliation{National Astronomical Observatory of Japan, 2-21-1 Osawa, Mitaka, Tokyo 181-8588, Japan}
\affiliation{Department of Astronomical Science, The Graduate University for Advanced Studies, SOKENDAI, 2-21-1 Osawa, Mitaka, Tokyo 181-8588, Japan}

\author[0000-0002-2026-8157]{Kenji Furuya}
\affiliation{National Astronomical Observatory of Japan, 2-21-1 Osawa, Mitaka, Tokyo 181-8588, Japan}

\author[0000-0003-1534-5186]{Richard Teague}
\affiliation{Department of Earth, Atmospheric, and Planetary Sciences, Massachusetts Institute of Technology, Cambridge, MA 02139, USA}

\author[0000-0003-1413-1776]{Charles J. Law}
\altaffiliation{NASA Hubble Fellowship Program Sagan Fellow}
\affiliation{Department of Astronomy, University of Virginia, Charlottesville, VA 22904, USA}

\author[0000-0002-6034-2892]{Takashi Tsukagoshi}
\affiliation{{  Faculty of Engineering, Ashikaga University, Ohmae-cho 268-1, Ashikaga, Tochigi, 326-8558, Japan}}

\author[0000-0002-0226-9295]{Seokho Lee}
\affiliation{Korea Astronomy and Space Science Institute (KASI), 776 Daedeokdae-ro, Yuseong-gu, Daejeon 34055, Republic of Korea}

\author[0000-0003-1817-6576]{Christian Rab}
\affiliation{University Observatory, Faculty of Physics, Ludwig-Maximilians-Universit\"at M\"unchen, Scheinerstr. 1, 81679 Munich, Germany}
\affiliation{Max-Planck-Institut f\"ur extraterrestrische Physik, Giessenbachstrasse 1, 85748 Garching, Germany}

\author[0000-0001-8798-1347]{Karin I. \"{O}berg}
\affiliation{Center for Astrophysics \textbar Harvard \& Smithsonian, 60\ Garden St., Cambridge, MA 02138, USA}

\author[0000-0002-8932-1219]{Ryan A. Loomis}
\affiliation{National Radio Astronomy Observatory, 520 Edgemont Rd., Charlottesville, VA 22903, USA}



\begin{abstract}
Recent measurements of carbon isotope ratios in both protoplanetary disks and exoplanet atmospheres have suggested a possible transfer of significant carbon isotope fractionation from disks to planets.
For a clearer understanding of the isotopic link between disks and planets, it is important to measure the carbon isotope ratios in various species.
In this paper, we present a detection of the \jcn\ $N=2-1$ hyperfine lines in the TW Hya disk with the Atacama Large Millimeter/submillimeter Array.
This is the first spatially-resolved detection of \jcn\ in disks, which enables us to measure the spatially resolved \icn/\jcn\ ratio for the first time.
We conducted non-local thermal equilibrium modeling of the \jcn\ lines in conjunction with previously observed \icn\ lines to derive the kinetic temperature, ${\rm H_2}$ volume density, and column densities of \icn\ and \jcn.
The ${\rm H_2}$ volume density is found to range between $ (4 - 10)\times10^7 \ {\rm cm^{-3}}$, suggesting that CN molecules mainly reside in the disk {\rm upper layer}.
The \icn/\jcn\ ratio is measured to be $ 70^{+9}_{-6}$ at $30 < r < 80$ au from the central star, which is similar to the $\rm ^{12}C/^{13}C$ ratio in the interstellar medium.
However, this value differs from the previously reported values found for other carbon-bearing molecules (CO and HCN) in the TW Hya disk.
This could be self-consistently explained by different emission layer heights for different molecules combined with preferential sequestration of $\rm ^{12}C$ into the solid phase towards the disk midplane.
This study reveals the complexity of the carbon isotope fractionation operating in disks.
\end{abstract}



\section{Introduction} \label{sec:intro}

The physical and chemical structure of protoplanetary disks are now routinely well-studied using the unique capabilities of the Atacama Large Millimeter/submillimeter Array (ALMA).
It is now clear that the rings and gaps are ubiquitous in the mm dust \citep[e.g., the DSHARP project,][]{andr18, huan18_dsharp} and molecular gas distributions \citep[e.g., the MAPS project,][]{ober21maps, law21b} of large planet-forming disks.
At least some of these sub-structures are thought to be related to ongoing planet formation in these disks.
The next question that we should focus on is {\it what kind of disk environment forms what kind of planetary system?}
The link between disks and exoplanetary systems as well as our solar system is still not understood well.
To link disks to the composition of planets, chemical information is vital.

For example, \citet{ober11} proposed the carbon-to-oxygen elemental ratio (C/O) as a tracer of the birthplace of gas giant planets.
Since the main carriers of carbon and oxygen, CO and H$_2$O, have different condensation temperatures, their snow lines are located at different distances from the central star in a disk.
As a consequence, the gas-phase C/O ratio is expected to dramatically change as a function of radius.
Therefore, it is predicted that planetary cores at different radii may accrete the disk gas with different C/O.

Similarly, isotopologue ratios are powerful material tracers in general.
In this paper, we target the carbon isotope ratio, $\rm ^{12}C/^{13}C$.
Since carbon is the fourth most abundant element in the universe, and, one of the main ingredients of planetary bodies and ultimately organic molecules, it is essential to understand its behavior including the isotope chemistry during planet formation processes.
In the local interstellar medium (ISM), $\rm ^{12}C/^{13}C$ is measured to be $\sim 69$ \citep{wils99}.
Despite its importance, the measurement of $\rm ^{12}C/^{13}C$ is quite sparse in both protoplanetary disks and planetary bodies.
In the solar system comets, it is known that the $\rm ^{12}C/^{13}C$ ratio is almost constant at $91.0\pm3.6$, however, the available heliocentric radii are limited \citep{mumm11}.
There are a few measurements on exoplanetary atmospheres.
For instance, \citet{yzha21a} observed a hot Jupiter and found a low $\rm ^{12}CO/^{13}CO$ ratio of $\sim 30$.
On the other hand, \citet{line21} suggested that the $\rm ^{12}CO/^{13}CO$ ratio in a brown dwarf is enhanced compared to the ISM value.

Among protoplanetary disks, few sources have been as extensively studies as the TW Hya disk, which is the nearest disk to Earth \citep[$D\sim 60$ pc;][]{gaia16, gaia21}.
\citet{zhan17} and \citet{yosh22} found a significantly low $\rm ^{12}CO/^{13}CO$ of $\sim 20-40$ at a similar radius from the central star.
On the other hand, \citet{hily19} showed that the $\rm ^{12}C/^{13}C$ ratio in HCN is even higher than the ISM value.
To link the environment of disks and planets, we need to better understand the origins of such differences in measured carbon isotope ratios as well as map them spatially across the disks both radially and vertically.
This requires more observational work. 

In this paper, we report a measurement of the \rcn\ ratio using ALMA archival observations.
In the next section, we describe the details of the archival observations.
This is the first spatially-resolved detection of \jcn\ in a disk.
We use a non-LTE analysis to derive the \rcn\ ratio and show results in Section \ref{sec:ana}.
We discuss the results in Section \ref{sec:disc}, and summarize this paper in Section \ref{sec:sum}.

\section{Observations} \label{sec:obs}

We analyzed ALMA archival observations of the $\rm ^{13}CN$ and $\rm ^{12}CN$ lines in the TW Hya disk to derive the \rcn\ ratio.
The following data reduction and imaging was performed with Common Astronomy Software Applications \citep[CASA;][]{mcmu07} modular version 6.2.5 with the exception of the pipeline calibration which was performed in the CASA version used for the original publication.
All images we used for analysis are convolved by an elliptical Gaussian to match the final beam to a circular Gaussian with an FWHM of $0\farcs5$ using the CASA tasks \texttt{ imsmooth}, which is the smallest common resolution we can achieve.
\begin{table*}
\centering
%
\begin{threeparttable}[hbtp]
\caption{Observation details}
\label{tab:obs}
\begin{tabularx}{\linewidth}{cccccccc}
\midrule
Transition & Representative Freq. & Project ID & {\tt robust} & Original Beam\tnote{a} & Chan. Wid. & RMS\tnote{b} &  Int. Time \\
\ & (GHz) & \ & \ & \ & (${\rm km\ s^{-1}}$) & (${\rm mJy\ beam^{-1}}$) & ($\rm min$)   \\
\midrule
${\rm ^{13}CN\ N=2-1}$ & 217.467150 & 2016.1.01375.S & 0.0 & $0\farcs46 \times 0\farcs44, -88^\circ$ & 0.34 & 1.0 & 41 \\
${\rm ^{12}CN\ N=2-1}$ & 226.874781 & 2018.A.00021.S & 1.0 & $0\farcs42 \times 0\farcs33, 88^\circ$ & 0.10 & 0.89 & 228 \\
${\rm ^{12}CN\ N=1-0}$ & 113.490970 & 2017.1.01199.S & 0.0 & $0\farcs50 \times 0\farcs45, 42^\circ$ & 0.10 & 3.1 & 105 \\
\midrule
\end{tabularx}
\begin{tablenotes}
\item{a} The original CLEAN images were convolved a Gaussian to make the final beam to $0\farcs5\times0\farcs5$. 
\item{b} Measured after matching the beam
\end{tablenotes}

\end{threeparttable}
\end{table*}
We summarize the observation details in Table \ref{tab:obs} and describe in the following.

\subsection{\jcn}
The ALMA archival data  (ID: 2016.1.01375.S, PI: C. Rab) consists of two execution blocks (EBs).
The first EB was observed on Nov.19, 2016 with 43 antennas, while the second EB was observed on Nov.20, 2016 with 42 antennas.
For both EBs, the integration time was 41 minutes, and the baseline range was 15-704 m.
For both EBs, the quasar J1037-2934 was used as the gain calibrator while J1107-4449 was used as a flux calibrator.
As the bandpass calibrator, J1037-2934 and J1107-4449 were used for the first and second EBs, respectively.
The \jcn\ $N=2-1$ lines were observed in the spectral window 1 of the baseband 1 with the spectral resolution of $\sim 0.67\ {\rm km\ s^{-1}}$ .
Table.\ref{tab:13cnlines} shows a full listing of the hyperfine lines that are covered in the archival data.

For data reduction, we first executed the pipeline calibration by running \texttt{scriptForPI.py}. 
Then, we CLEANed the continuum SPW using the CASA task \texttt{tclean} after flagging the detected emission lines.
We performed five rounds of phase self-calibration with solution intervals of the duration of the EBs, 1200s, 600s, 300s, and 100s as well as one round of amplitude self-calibration with the duration of the EBs.
The solutions were applied to the SPW containing \jcn\ $N=2-1$ lines without flagging lines, and the continuum emission was subtracted by fitting a linear function to line-free channels on the visibility space using the \texttt{uvcontsub} task.
Finally, the velocity channels that contain the \jcn\ lines were imaged with a channel spacing of 0.34 $\rm km\ s^{-1}$ (half of the velocity resolution) and robust parameter of 0.5.
For the CLEAN mask, a Keplerian mask made with a Python script \texttt{Keplerian\_mask} \citep{kepmask} was used assuming the disk inclination angle of $5.8^\circ$, position angle of $151.6^\circ$, stellar mass of $0.81\ M_\odot$, the systemic velocity of $2.84\ \rm km\ s^{-1}$, following \citet{teag19}.
The outer radius of the mask was set to 3.0 arcsec to contain all emissions in the mask.
The CLEAN image cube was then convolved by an elliptical Gaussian to match the final beam to a circular Gaussian with an FWHM of $0\farcs5$ using \texttt{ imsmooth }, and primary beam correction was applied with \texttt{impbcor}.
The root mean square (RMS) noise level is estimated to be 1.0 mJy, which is calculated by averaging the RMS value of the first and last three line-free channels of the image cube before primary beam correction.

\subsection{\icn}
We also analyzed ALMA observations of the ${\rm ^{12}CN}\ N=2-1$ lines  (ID:2018.A.00021.S, PI: R. Teague).
The measurement sets include ${\rm CO}\ J=2-1$ and ${\rm CS}\ J=3-2$ lines as well.
Details about these observations are described in \citet{teag22}.
After downloading the data and running the pipeline calibration, we first imaged and self-calibrated the continuum only using the short baseline data.
Then, the data was combined with the longer baseline EBs.
We noticed that the amplitude as a function of the uv distance of one EB systemically deviates from others, therefore, that EB is totally flagged.
We performed six rounds of phase self-calibration to the combined data.
The solution interval is decreased from the duration of EBs to 30 seconds.
Then, we ran one round of amplitude self-calibration with the duration of EBs.
The solutions of the self-calibration were applied to the spectral windows containing the \icn\ lines.
After subtracting the continuum emission, two SPWs that contain the \icn\ lines were CLEANed using \texttt{tclean} with a velocity channel width of $0.1\ {\rm km\ s^{-1}}$ and robust parameter of 1.0 using a CLEAN mask generated by a Python script, \texttt{Keplerian\_mask}.

We also used the $\rm ^{12}CN\ N=1-0$ data from the archive (ID: 2017.1.01199.S, PI: R. Loomis).
The details of these observations were described in \citet{teag20}.
We ran pipeline calibration and self-calibration in a similar way of \citet{teag20} and CLEANed with a robust parameter of 0.
Both of the \icn\ data cubes were also re-convolved by a Gaussian that matches the final beam to the $0\farcs5$ circular Gaussian for consistency. 

We also checked the absolute flux consistency among the three observation sets.
We extracted a spectral energy distribution (SED) from the peak intensity of continuum images corresponding to the \icn\ $N=1-0,\ 2-1$ and \jcn\ $N=2-1$ image cubes.
The continuum images are also re-convolved and the beam is a circular Gaussian with a FWHM of $0\farcs5$.
The SED can be well fitted by a power-law curve with an index of 2.13, which is consistent with the literature \citep{tsuk16, ueda20, maci21, tsuk22}.
The deviations from the best-fit curve are $\sim -0.3\%,\ 3.1\%$, and $-2.7\%$ for the continuum images corresponding to the \icn\ $N=1-0,\ 2-1$ and \jcn\ $N=2-1$ images.
Since these deviations are sufficiently small, we neglect the uncertainty on the absolute flux scaling in the following analysis.

\begin{table*}
\centering
\begin{threeparttable}[hbtp]
\caption{Spectroscopic properties of the \jcn\ hyperfine $N=2-1$ lines observed in the data.}
\label{tab:13cnlines}

\begin{tabular}{ccccccc}
\midrule
$J$ & $F_1$ & $F$ & Frequency [GHz] & $A\ [{\rm s^{-1}}]$ &  $g_u$  & $E_{\rm up}\ {\rm [K]}$  \\
\midrule
3/2-1/2 &     1-0 &   0-1 &      217.264639 &  $6.05 \times 10^{-4}$ &    1 &                $15.7$ \\
 3/2-1/2 &     2-1 &   1-2 &      217.276438 &  $2.70 \times 10^{-5}$ &    3 &                $15.6$ \\
 3/2-1/2 &     1-0 &   1-1 &      217.277680 &  $5.80 \times 10^{-4}$ &    3 &                $15.7$ \\
 3/2-1/2 &     2-1 &   2-2 &      217.286804 &  $2.42 \times 10^{-4}$ &    5 &                $15.6$ \\
 5/2-3/2 &     2-2 &   2-1 &      217.287543 &  $3.08 \times 10^{-5}$ &    5 &                $15.7$ \\
 5/2-3/2 &     2-2 &   3-2 &      217.289801 &  $2.22 \times 10^{-5}$ &    7 &                $15.7$ \\
 3/2-1/2 &     2-1 &   1-1 &      217.290823 &  $4.00 \times 10^{-4}$ &    3 &                $15.6$ \\
 5/2-3/2 &     2-2 &   1-1 &      217.294470 &  $1.64 \times 10^{-4}$ &    3 &                $15.7$ \\
 3/2-1/2 &     2-1 &   1-0 &      217.296605 &  $5.30 \times 10^{-4}$ &    3 &                $15.6$ \\
 5/2-3/2 &     2-2 &   2-2 &      217.298937 &  $1.50 \times 10^{-4}$ &    5 &                $15.7$ \\
 3/2-1/2 &     2-1 &   2-1 &      217.301175 &  $7.16 \times 10^{-4}$ &    5 &                $15.6$ \\
 3/2-1/2 &     2-1 &   3-2 &      217.303191 &  $9.60 \times 10^{-4}$ &    7 &                $15.6$ \\
 3/2-1/2 &     1-0 &   2-1 &      217.304927 &  $5.31 \times 10^{-4}$ &    5 &                $15.7$ \\
 5/2-3/2 &     2-2 &   1-2 &      217.305904 &  $5.25 \times 10^{-5}$ &    3 &                $15.7$ \\
 5/2-3/2 &     2-2 &   3-3 &      217.306117 &  $1.88 \times 10^{-4}$ &    7 &                $15.7$ \\
 5/2-3/2 &     2-2 &   2-3 &      217.315147 &  $3.16 \times 10^{-5}$ &    5 &                $15.7$ \\
 5/2-3/2 &     2-1 &   3-2 &      217.428563 &  $7.67 \times 10^{-4}$ &    7 &                $15.7$ \\
 5/2-3/2 &     2-1 &   2-1 &      217.436350 &  $5.89 \times 10^{-4}$ &    5 &                $15.7$ \\
 5/2-3/2 &     2-1 &   2-2 &      217.437702 &  $1.76 \times 10^{-4}$ &    5 &                $15.7$ \\
 5/2-3/2 &     2-1 &   1-1 &      217.443722 &  $3.09 \times 10^{-4}$ &    3 &                $15.7$ \\
 5/2-3/2 &     2-1 &   1-0 &      217.443722 &  $4.35 \times 10^{-4}$ &    3 &                $15.7$ \\
 5/2-3/2 &     2-1 &   1-2 &      217.444666 &  $1.87 \times 10^{-5}$ &    3 &                $15.7$ \\
 5/2-3/2 &     3-2 &   4-3 &      217.467150 &  $1.01 \times 10^{-3}$ &    9 &                $15.7$ \\
 5/2-3/2 &     3-2 &   3-2 &      217.467150 &  $8.92 \times 10^{-4}$ &    7 &                $15.7$ \\
 5/2-3/2 &     3-2 &   2-1 &      217.469151 &  $8.43 \times 10^{-4}$ &    5 &                $15.7$ \\
 5/2-3/2 &     3-2 &   2-2 &      217.480559 &  $1.59 \times 10^{-4}$ &    5 &                $15.7$ \\
 5/2-3/2 &     3-2 &   3-3 &      217.483606 &  $1.14 \times 10^{-4}$ &    7 &                $15.7$ \\
 5/2-3/2 &     3-2 &   2-3 &      217.496736 &  $4.70 \times 10^{-6}$ &    5 &                $15.7$ \\
\midrule
\end{tabular}
\begin{tablenotes}
\item Notes. $J, F_1$ and $F$ are the quantum numbers. $A$ and $g_u$ are the Einstein A coefficients and upper state degeneracy. The data is taken from the CDMS database \citep{mull01, mull05, endr16}. 
\end{tablenotes}

\end{threeparttable}
\end{table*}

\section{Analysis and Results}\label{sec:ana}

\subsection{The image of \jcn\ and matched filter analysis}

\begin{figure*}[hbtp]
    \epsscale{1.2}
    \plotone{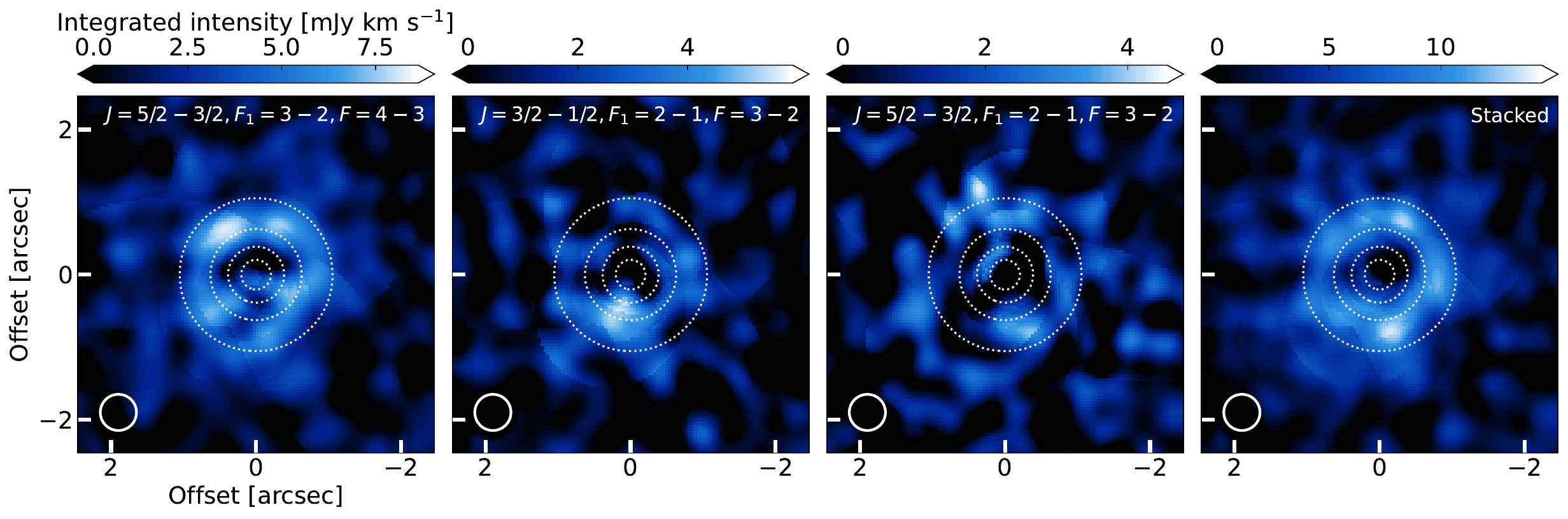}
    \caption{ Moment zero maps of the three brightest \jcn\ lines and the stacked moment zero maps of the three lines. The white dotted contours represent the 217 GHz continuum emission, starting at the brightness temperature of 1K and drawn every 3K. The synthesized beam is shown in the lower-left corner of each panel.}
    \label{fig:mom0}
\end{figure*}

We created moment zero maps of the three brightest lines with the same Keplerian mask used in the CLEANing process using the CASA task \texttt{immoments} and then stacked each of these moment zero maps to maximize the signal-to-noise ratio.
Figure \ref{fig:mom0} shows the resulting \jcn\ $N=2-1$ moment maps.
The \jcn\ $N=2-1$ emission exhibits a ring-like morphology, which is similar to the \icn\ lines \citep{teag20}.
Also, the emission is extended beyond the continuum disk that is shown in the white dotted contour.

\begin{figure}[hbtp]
    \epsscale{1.2}
    \plotone{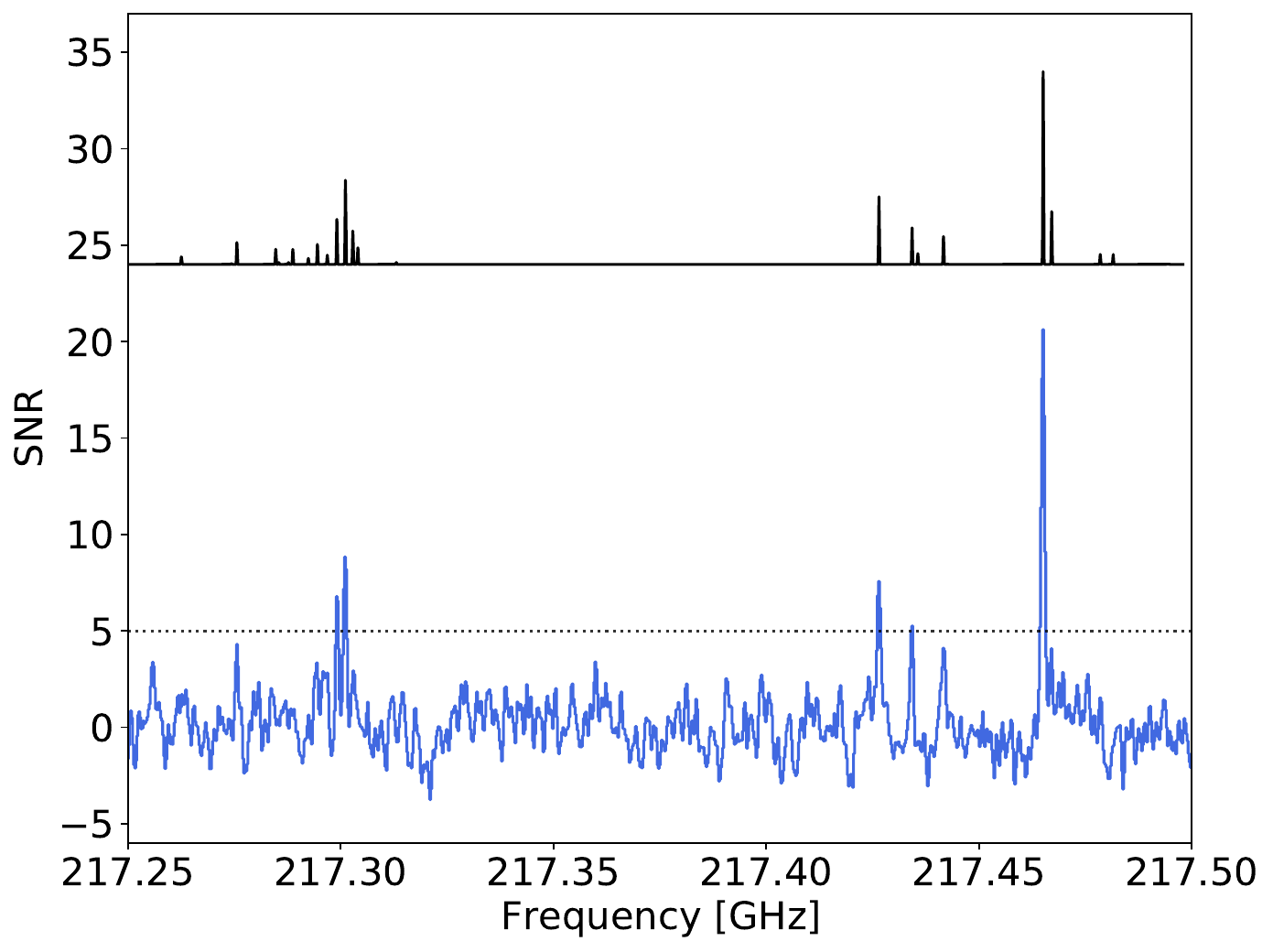}
    \caption{ Matched filter response at 217.25--217.50 GHz. The black line shows the model relative optical depth of the \jcn\ lines at $T=40$ K. All the lines shown here are from \jcn.}
    \label{fig:mf}
\end{figure}

Although the \jcn\ lines are clearly detected on the image plane, we also ran the matched filter analysis \citep{loom18}.
The Keplerian mask was used to generate the filter kernel.
The mask is similar to that used for CLEANing of the \jcn\ lines but with a smaller outer radius of 1.6 arcsec just to contain the emission of the brightest line (${\rm ^{13}CN}\ N=2-1\ J=5/2-3/2\ F_1=3-2\ F=4-3$).
The cross-correlation between the observed visibilities and filter kernel was calculated by a Python script \texttt{VISIBLE} \citep{loom18}.
The response spectrum is shown in Figure \ref{fig:mf}. 
Assuming the local thermal equilibrium (LTE), a predicted relative optical depth spectrum at $T=40$ K suggested by \citet{teag20} is also plotted.
Five hyperfine lines are detected in a significance of $>5\sigma$, and the brightest line, ${\rm ^{13}CN}\ N=2-1\ J=5/2-3/2\ F_1=3-2\ F=4-3$, exceeds a $\sim 20 \sigma$ significance.
This is the first spatially-resolved detection of \jcn\ in a protoplanetary disk.
Note that \citet{phuo21} reported a detection of \jcn\ in the GG Tau A disk with $\sim 6-8\sigma$ significance.
The MAPS project \citep{ober21maps} also observed the ${\rm ^{13}CN}\ N=2-1$ line but it was not detected.
This is because they only covered the first seven hyperfine lines listed in Table \ref{tab:13cnlines}, whose Einstein A coefficients (times upper state degeneracy) are relatively smaller.

\subsection{Non-LTE modeling of azimuthally averaged spectra of \icn\ and \jcn\ }
\begin{figure*}[hbtp]
    \epsscale{1.1}
    \plotone{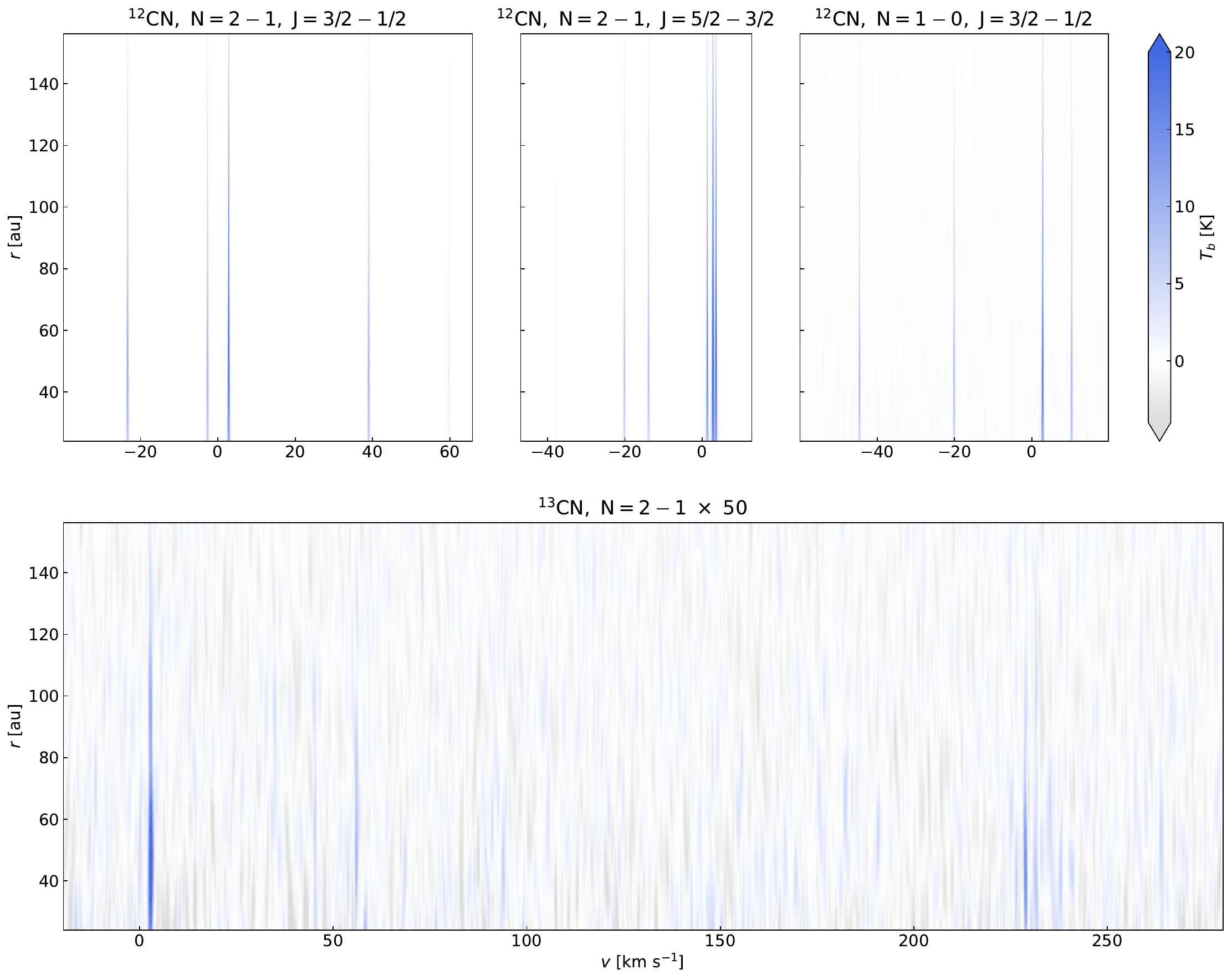}
    \caption{ Azimuthally stacked CN spectra. The horizontal and vertical axes indicate the velocity with respect to the rest frame of the brightest line and radius from the star, respectively.}
    \label{fig:gf}
\end{figure*}

\begin{figure}[hbtp]
    \epsscale{1.1}
    \plotone{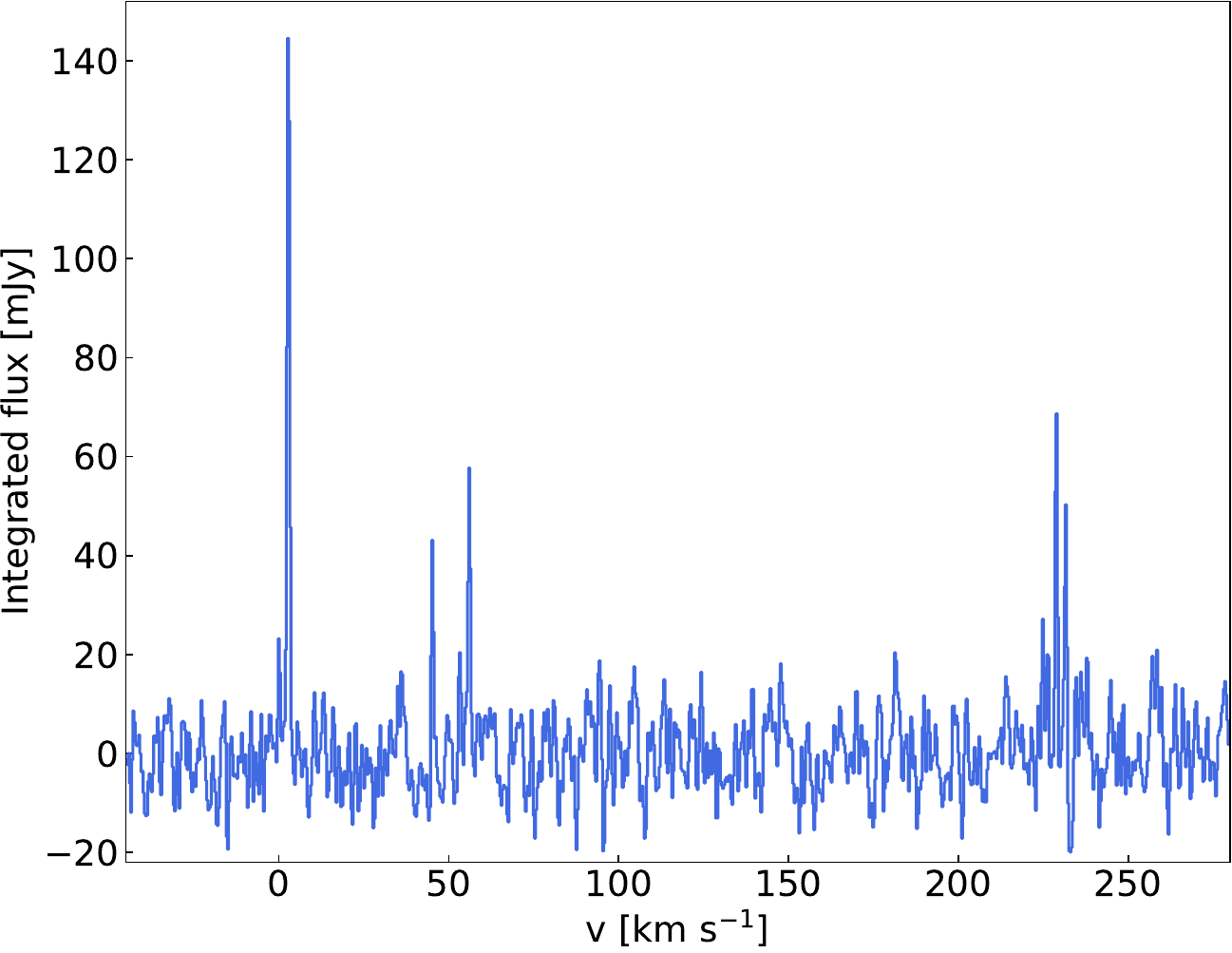}
    \caption{ Disk integrated flux of the \jcn\ lines after correcting for the Keplerian rotation. The horizontal and vertical axes indicate the velocity with respect to the rest frame of the brightest line and flux density, respectively.}
    \label{fig:integrated}
\end{figure}

For further analysis, the spectra at each position were azimuthally stacked assuming Keplerian rotation using the python package {\tt gofish} \citep{teag16}.
We used the same disk geometric parameters as used for the CLEAN mask.
We also specified the disk center position by fitting a 2D Gaussian to the corresponding continuum images. 
The resultant stacked spectra of the \icn\ and \jcn\ lines as a function of the radius from the central star and velocity shift from the rest frame of the brightest lines in each hyperfine group are presented in Figure \ref{fig:gf}.
For the \jcn\ lines, we also present the disk integrated flux spectrum after stacking with {\tt gofish} in Figure \ref{fig:integrated}.
Here, we adopted the disk outer radius of 3.0 arcsec.

To derive the physical quantities, we generated synthetic spectra and compared them with the azimuthally stacked spectra.
\citet{teag20} previously found that the \icn\ lines are not in LTE by comparing independently derived kinetic and excitation temperatures.
Therefore, we performed non-LTE modeling.
The peak optical depth $\tau_{c, i}$ and excitation temperature $T_{{\rm ex}, i}$ of each hyperfine line $i$ are calculated in three-dimensional grids for each SPW by using a Python wrapper of the non-LTE radiative transfer code RADEX \citep{vand07}, \texttt{SpectralRadex}\footnote{\url{https://github.com/uclchem/SpectralRadex}} \citep{hold21}.
We adopted publically available spectroscopic data \citep{klis95, thom68, kalu15} via the LAMDA database \citep{scho05} for the \icn\ lines.
{For the \jcn\ lines, we used the LAMDA format file in the EMAA database \footnote{\url{https://emaa.osug.fr}} which is based on \citet{mull05} and \citet{flow15}.}
Note that the rate coefficients are available only for collision with para-$\rm H_2$, so we ignored ortho-$\rm H_2$ in our analysis. This may not be an issue if the rate coefficients against para- and ortho-$\rm H_2$ are similar as seen in CO since the ortho-to-para ratio of $\rm H_2$ at $30-50$ K is $0.031-0.30$ under the thermal equilibrium \citep{flow84, flow85} which is expected in disks \citep{aika18}.
However, as in the case of \icn, these rate coefficients could be significantly different.
New spectroscopic data for collision with ortho-$\rm H_2$ are required to model the \jcn\ lines more robustly.

Each RADEX grid consists of the kinetic temperature $T_k$, column density of \icn\ or \jcn\,  $N({\rm ^{12}CN}), N({\rm ^{13}CN})$, and ${\rm H_2}$ volume density $n({\rm H_2})$.
Here, we assumed that all lines share the same $T_k$ and $n({\rm H_2})$, and the line width is identical to the thermal width.
We also adopted the background emission of the black body radiation at 2.73 K.
Note that the dust continuum emission is similar to or fainter than 2.73 K at $100-200$ GHz at $R>40$ au.
While $T_{\rm kin}$ is gridded linearly, $N$ and $n({\rm H_2})$ are sampled logarithmically.
The parameter ranges are listed in Table \ref{tab:prior}.
\begin{table}[hbtp]
  \caption{Parameter ranges in our fitting}
  \label{tab:prior}
  \centering
  \begin{tabular}{ccc}
    \hline
    Parameters  & Ranges & Units\\
    \hline \hline
    $\log_{10} n({\rm H_2})$  & $[6, 12]$  & $\rm cm^{-3}$ \\
    $\log_{10} N({\rm ^{12}CN}), \log_{10} N({\rm ^{13}CN})$  & $[11, 16]$  & $\rm cm^{-2}$ \\
    $T_k$  & $[10, 100]$  & $\rm K$ \\
    $v_{\rm sys}$  & $[2.6, 3.0]$  & $\rm km\ s^{-1}$ \\
    $\Delta v_{j}$  & $[0.02, 2.0]$  & $\rm km\ s^{-1}$ \\
    \hline
  \end{tabular}
\end{table}
The number of the grid is 100 along each dimension.

Then, we generated synthetic spectra of all observed \icn\ and \jcn\ lines at each radius using $\tau_{c, i}$ and $T_{{\rm ex}, i}$ calculated by {\tt SpectralRadex} and fitted them to the observed spectra as the following.
First, $T_k$, $n({\rm H_2})$, and $N({\rm ^{12}CN})$ (or $N({\rm ^{13}CN})$) are inputted to the model grids, and they return $\tau_{c, i}$ and $T_{{\rm ex}, i}$ for each hyperfine line $i$ by interpolation of the grids.
Then, we calculated the optical depth profile $\tau_i(v)$ of each line by adopting thermal line width, where $v$ is the velocity with respect to the brightest line in each group.
We assumed that the molecular and dust emitting regions are vertically distinct; while the dust grains are settled on the midplane, the molecular emission originates from higher altitude regions.
The intensity coming from one-sided molecular layer $I_{\rm mol}$ is expressed as
\begin{equation}
I_{\rm mol}(v) = \displaystyle \frac{\sum_i B(T_{{\rm ex}, i}) \displaystyle  \frac{\tau_i(v)}{2}  }{ \sum_i \displaystyle 
 \frac{\tau_i(v)}{2}   } \left\{ 1 - \exp\left( -\sum_i \displaystyle \frac{\tau_i(v)}{2}  \right) \right\},
\end{equation}
following \citet{hsie15}, where $B(T_{{\rm ex}, i})$ is the Planck function at the temperature $T_{{\rm ex}, i}$ at the corresponding frequency.
This formulation can treat the line overlapping approximately.
{Since we consider the three-layer (molecules in the backside, dust grains at the midplane, and molecules in the frontside) model along the line of sight, the resultant intensity $I_{\rm mod}$ after subtracting the continuum emission can be written as }
\begin{equation}
I_{\rm mod}(v) = \left\{  I_{\rm mol} \exp(-\tau_d) + I_c \right\} \exp\left( -\sum_i \displaystyle \frac{\tau_i(v)}{2}  \right)  + I_{\rm mol} - I_c,
\end{equation}
where $\tau_d$ and $I_c$ are the dust optical depth at each wavelength taken from \citet{maci21} and observed continuum emission, respectively.
{Note that the factors $\exp{(-\tau_d)}$ and $\exp({-\sum_i \tau_i(v)/2})$ express the continuum absorption at the midplane and molecular line absorption in the frontside, respectively.  }
The model spectrum $I_{\rm mod}(v)$ is generated with a $15$ kHz interval, shifted by the systemic velocity $v_{\rm sys}$, and convolved by a Gaussian with {  a standard deviation} of $\Delta v_j$ in each SPW ($j=1-4$), approximating any instrumental broadening effect due to the velocity resolution of the correlator and Keplerian shear in a beam \citep[e.g.,][]{berg21, muno23}.
As a result, we have nine free parameters at each radius, $n({\rm H_2}), T_k, N({\rm ^{12}CN}), N({\rm ^{13}CN}), v_{\rm sys}$ and $\Delta v_j$ for four SPWs.
The generated synthetic spectra are compared with the observed spectra shown in Figure \ref{fig:gf} in terms of the log-likelihood function
\begin{equation}
\log{ \mathcal{L} } = -\frac{1}{2} \sum_v \left( \frac{I_{\rm mod}(v) - I_{\rm obs}(v) }{ \sigma_{\rm obs} } \right)^2,
\end{equation}
where the summation is done over all { velocity channels of the observations}. 
$I_{\rm obs}(v)$ and $\sigma_{\rm obs}$ are the azimuthally stacked spectra and their standard deviation, respectively.
$\sigma_{\rm obs}$ is calculated for independent velocity bins; for the \icn\ lines, since the original velocity resolution is narrower than the final channel width, the calculated standard deviations are directly used.
On the other hand, the standard deviation calculated in the \jcn\ spectra is multiplied by $\sqrt{2}$ because the channel width is half of the velocity resolution.
In practice, with the uniform prior shown in Table \ref{tab:prior}, we used \texttt{emcee} \citep{fore13} and sampled the posterior distribution by the Markov Chain Monte Carlo (MCMC) method.
We set our MCMC chains with 64 walkers and 10,000 steps and a burn-in of the first 6,000 steps.
The marginal posterior distributions are almost Gaussian for all parameters and radii.
We present the representative corner plots in Appendix \ref{appendix}.

\subsection{Fitting Results}
\begin{figure*}[hbtp]
    \epsscale{1.1}
    \plotone{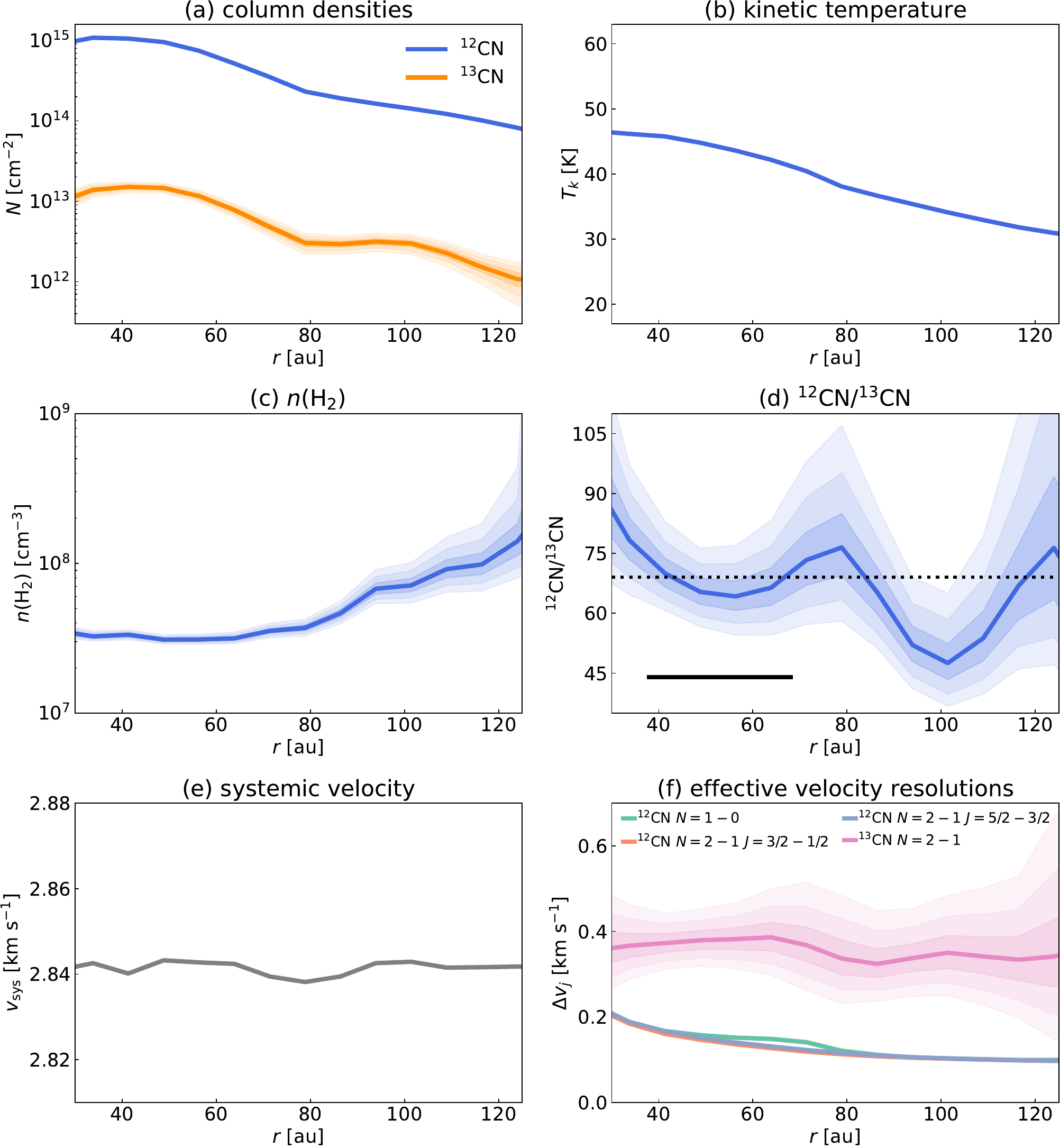}
    \caption{ MCMC fitting results. (a) column densities of \icn\ and \jcn (b) kinetic temperature $T_k$. (c) $\rm H_2$ volume density $n({\rm H_2})$. (d) the \icn/\jcn\ ratio with black dashed line and bar, indicating the ISM carbon isotope ratio and beam size, respectively. (e) systemic velocity. (f) effective velocity resolutions for each SPW. In all panels, the shading represents the 1,2 and 3-sigma confidence contours estimated from the posterior distributions. Uncertainties in a few parameters are smaller than the line thickness.}
    \label{fig:results}
\end{figure*}

We plot the fitting results in Figure \ref{fig:results}.
Panel (a) shows the column density profiles of \icn\ and \jcn. The column density of \icn\ ranges $10^{14-15}\ {\rm cm^{-2}}$ at r = $30-120$ au.
This value is broadly consistent with the disk averaged $\rm ^{12}CN$ column density of $(9.6\pm1.0) \times 10^{13}\ {\rm cm^{-2}}$ obtained from the data observed by the Atacama Pathfinder Experiment telescope \citep{kast14} {and that of $\sim 10^{14} {\rm cm^{-2}}$ estimated by \citet{hily17} using ALMA observations. }
The slope of the column density profiles changes at $r\sim 80$ au, corresponding to the truncation radius of the mm-continuum disk \citep[e.g.,][]{ilee22}.

The best-fit kinetic temperature (panel b) is lower than that derived from the brightness temperature of ${\rm ^{12}CO}\ J=3-2$ \citep{huan18} but higher than ${\rm ^{13}CO}\ J=3-2$ \citep{nomu21}, suggesting that CN molecules exist at a higher altitude than most of CO molecules.
This is also evident in the $n({\rm H_2})$ profile (panel c), which shows that $n({\rm H_2})$ at $r < 80$ au is $\sim 4\times10^7\ {\rm  cm^{-3}}$.
The gas surface density and gas scale height at $r=60$ au are $\sim 4\ {\rm g\ cm^{-2}}$ and $\sim5$ au, respectively, in the fiducial TW Hya disk model of \citet{cala21}.
Assuming these values, a mean molecular weight of 2.37, and that vertical gas distribution follows a Gaussian, $n({\rm H_2}) = 4\times10^7\ {\rm cm^{-3}}$ corresponds to an altitude of $z/r\sim0.27$ in the model.
Meanwhile, $n({\rm H_2})$ is gradually increasing toward the outer region and reaches $\sim 10^8\ {\rm cm^{-3}}$ at 100 au, which corresponds to $z/r\sim0.22$ in the fiducial model of \citet{cala21}.
The derived $n({\rm H_2})$ are roughly consistent with those found by \citet{muno23} from modeling of spatially-resolved multi-line DCN and DCO$^+$ observations.
The derived kinetic temperature and $\rm H_2$ volume density are consistent with the previous analysis by \citet{teag20} {as well as previous disk models \citep{clee15, lee21, nomu21}.}

We also derive the \icn/\jcn\ column density ratio (panel d).
The ratio is consistent with the ISM value \citep[$\sim 69; $][]{wils99} over $30 < r < 80$ au.
We calculated the median and [16, 84] th percentiles of posteriors, and find \icn/\jcn\ $\simeq 70^{+9}_{-6}$ is a representative value.
Also, it is likely that there is a dip of \icn/\jcn\ at $r \sim 100$ au, although its deviation from the ISM value is only $\sim 3 \sigma$.
At the local minimum, \icn/\jcn\ is estimated to be $ 48^{+5}_{-4}$.

We also plot the best-fit systemic velocity and effective velocity resolutions for each SPW in panel (e) and (f).
The stable systemic velocities as a function of radius verify that the shift-and-stack process was successful.
The effective velocity resolutions for the \icn\ lines increase as the radius approaches the disk center. This is because the velocity gradient due to the Keplerian shear in one beam is larger near the disk center.

\section{Discussion}\label{sec:disc}

\subsection{Vertical distribution of CN}
The derived kinetic temperature and $\rm H_2$ volume density imply that the CN emission mainly arises from the {\rm upper layer of} the inner disk ($r \lesssim 80$ au) and relatively closer to the midplane in the outer disk ($r \gtrsim 80$ au).
The estimated altitude of the CN emitting region of $z/r\sim0.2-0.3$ is consistent with the previous study on TW Hya \citep{teag20} as well as the MAPS targets \citep{berg21}, but much lower than the Elias 2-27 disk \citep{pane22}.

It has been suggested that the vertical distribution of CN is strongly affected by the UV attenuation in a disk \citep[e.g,][]{aika06, cazz18}.
\citet{aika06} found that the location of the CN abundance peak in the vertical direction of the disks shifts toward the midplane when the UV attenuation by dust particles is weakened although the CN column density is not affected.
Indeed, the mm-continuum emission truncates around $r\sim80$ au \citep[e.g.,][]{ilee22}, and the infrared scattered light exhibits a gap at $r\sim80-100$ au \citep{boek17}, implying that the UV attenuation is weakened in the outer disk.
A recent chemical model dedicated of the TW Hya disk by \citet{nomu21} also shows that CN can exist near the midplane outside the dust disk, which is consistent with our results.
The MAPS project suggested that the CN/HCN column density ratio is correlated with the mm-dust continuum gaps although it is not one-to-one correspondence \citep{berg21}.
Our results showing enhancement of CN near the midplane outside the mm-continuum disk are in line with their results.

\subsection{ Carbon isotope fractionation scenarios in the TW Hya disk } \label{sec42}
The \icn/\jcn\ column density ratio between $30<r<80$ au is $70^{+9}_{-6}$, similar to the ISM ratio.
The carbon isotope ratio in the TW Hya disk has also been measured in other carbon-bearing molecules.
The carbon isotope fractionation pattern in the similar region of the TW Hya disk is quite complex.
A low $\rm ^{12}CO/^{13}CO$ ratio of $21\pm5$ is suggested by analyzing the $\rm ^{12}CO$ and ${\rm ^{13}CO}$ line wings at $r\sim70-110$ au \citep{yosh22}.
\citet{zhan17} also found a low $\rm ^{12}C^{18}O/^{13}C^{18}O$ ratio of $40^{+9}_{-6}$ at $r\sim21$ au.
On the other hand, \citet{hily19} measured the $\rm H^{12}CN/H^{13}CN$ ratio to be $86\pm4$ at $r\sim20-55$ au.
{Although we should be cautious about differences in methods, it is clear that different molecules have different carbon isotope ratios.}

To discuss carbon isotope fractionation in the TW Hya disk, it is necessary to estimate and compare column densities of carbon bearing molecules to understand which species is the major carbon carrier.
The column densities of CN and HCN are much lower than that of CO \citep{zhan17, lee21}.
The averaged column density of the neutral atomic carbon can be estimated to be $\sim 7 \times 10^{15}\ {\rm cm^{-2}}$ from the observed value in \citet{kama16a} using equations in \citet{oka01} and \citet{tsuk15} assuming LTE and excitation temperature of 50 K.
This value is also 1-2 orders of magnitude smaller than CO.
Hydrocarbons are other carbon bearing species in disks.
The column density of $\rm C_2H$ at $r=100$ au from the central star is $\sim 10^{13} - 10^{14}\ {\rm cm^{-2}}$ according to models by \citet{berg16}.
In summary, it is likely that the main carbon carrier in the TW Hya disk gas is CO.

The fact that CO is the main carbon carrier means that the bulk gas $\rm ^{12}C/^{13}C$ is significantly low, $\sim 20-40$.
In the gas of a protoplanetary disk, there are two mechanisms that can cause carbon isotope fractionation; isotope-selective photodissociation and isotope exchange reaction \citep[e.g.,][]{wood09, viss18}.
In both cases, however, the carbon isotope ratio in the bulk gas does not change.
Therefore, carbon isotope fractionation between the gas and solid phase is needed to explain the observed isotopologue ratios.
In other words, $\rm ^{12}C$ would be selectively incorporated in the solid phase.
Future observations with the James Webb Space Telescope could find evidence of $\rm ^{13}C$-poor solid material.

So, what did make CO in the gas phase $\rm ^{13}C$-rich?
It is suggested that the gas-phase CO is depleted within the timescale of a few Myr from the protostellar stage by observations and models ( e.g., \citealp{zhan20, krij20, furu22}, but cf. \citealp{pasc23}).
Simultaneously the gas-phase C/O ratio outside the CO snowline becomes even larger than unity \citep[e.g.,][]{ober11, krij20, bosm21b, cala23}.
Since the stellar age of TW Hya is estimated to be 3-10 Myr \citep{barr06, vacc11}, it is possible that the high C/O ($>1$) ratio has been sustained for more than a few Myr in the past.
In this case, the isotope exchange reaction 
\begin{equation}
\label{eq:CO}
{\rm ^{13}C^+ + ^{12}CO \rightleftharpoons ^{12}C^+ + ^{13}CO + 35\ K},
\end{equation}
can enhance the gas-phase $\rm ^{13}CO$ in a relatively low-temperature environment ($\lesssim 35$ K).
Subsequent chemical reactions and physical processes such as dust coagulation could lock the molecules produced from $\rm ^{12}C^+$ to the dust grains, making the bulk gas $\rm ^{13}C$-rich.
On the other hand, the carbon bearing molecules in the dust grains would be slightly $\rm ^{13}C$-poor compared to the ISM ratio.

Meanwhile, if CO is the main carbon carrier in the bulk gas, the current {  gas-phase} C/O ratio should be almost unity or less;
{for instance, if there are only CO molecules except for $\rm H_2$ in the gas phase, the abundance of carbon equals that of oxygen, or, $\rm C/O=1$. }
However, \citet{berg16} suggested that {  a high gas-phase C/O ratio of $>1$ is needed to reproduce the bright $\rm C_2H$ emission.}
This apparent contradiction might be solved by the vertical stratification of C/O as predicted by in some models \citep{krij20}.
The CO isotopologue ratio measured by \citet{yosh22} reflects the region near the CO snow surface, which is located deeper in the disk, as they used the optically thin line wings, while $\rm C_2H$ mainly reside in the upper layer where C/O exceeds unity.

Our analysis shows that CN traces higher altitude than $\rm ^{13}CO$, where the $\rm C_2H$ emission may arise {with $\rm C/O > 1$}.
{ The carbon isotopologue ratios in the upper layer could be modified by the in-situ reactions including reaction (\ref{eq:CO}).}
Therefore, the isotopologue ratios of CN and HCN may be explained in a way similar to the previous models \citep{wood09, viss18}.

Another scenario to explain the low carbon isotope ratio in CO, which is less likely but cannot be ruled out, is that the bulk gas- and solid-phase $\rm ^{12}C/^{13}C$ deviates from the local ISM ratio.
It is known that the $\rm ^{12}C/^{13}C$ in molecular clouds changes as a function of the distance from the Galactic center due to the chemical evolution of the galaxy.
However, observations of carbon bearing species in molecular clouds show some scatter even at the similar distance \citep{mila05}.
Therefore, we cannot exclude the possibility that the prestellar cloud of TW Hya was an outlier from the local isotope abundance.




\subsection{ CN and HCN isotopologue ratios}

The $\rm H^{12}CN/H^{13}CN$ ratio of $86\pm4$ \citep{hily19} is higher than \icn/\jcn\
of $70^{+9}_{-6}$.
\citet{lois20} listed several isotope exchange reactions related to HCN and CN including
\begin{eqnarray}
\label{eq:CN1} {\rm ^{13}C^+ + ^{12}CN} &\rightleftharpoons& {\rm ^{12}C^+ + ^{13}CN + 31.1\ K} \\
\label{eq:CN2} {\rm ^{13}C + ^{12}CN} &\rightleftharpoons& {\rm ^{12}C + ^{13}CN + 31.1\ K} \\
\label{eq:CN3} {\rm ^{13}C + H^{12}CN} &\rightleftharpoons& {\rm ^{12}C + H^{13}CN + 48.0\ K}.
\end{eqnarray}
Interestingly, reaction (\ref{eq:CN3}) has a larger zero point energy difference than reaction (\ref{eq:CN2}) with similar rate constants, which means that HCN should become more $\rm ^{13}C$-rich if there are only C, CN, and HCN.
{Indeed, in the TW Hya disk, the column density of the neutral carbon is comparable to or even larger than that of CN and HCN (Section \ref{sec42}).
However, our results showed that CN is more $\rm ^{13}C$-rich than HCN.
This implies that other reactions including reaction (\ref{eq:CN1}) may play an important role.}
More realistic models that consider both physical and chemical processes with a larger chemical reaction network are required to reveal the complex carbon isotope fractionation pattern { (Lee et al., submitted)}.

{We also note that a potential molecular stratification of CN and HCN (i.e. CN traces higher altitude than HCN) could also explain the isotopologue ratio that is inconsistent with the zero point energy differences.}


\subsection{ The \icn/\jcn\ dip at $r\sim100$ au }
It is more difficult to explain the origin of the low \icn/\jcn\ ratio of $48^{+5}_{-4}$ at $r\sim100$ au.
First, we can interpret this as a radial variation of \icn/\jcn. 
Indeed, in the \citet{viss18} model, it is shown that the column density ratios of $\rm H^{12}CN/H^{13}CN$ and $\rm ^{12}CN/^{13}CN$ have a radial dependence with a dip $\sim150-200$ au, although it is not a source-specific model for TW Hya.
In terms of a radial variation, \citet{yosh22} found a high value of $\rm ^{12}CO/^{13}CO > 84$ at $r\sim100$ au and they attributed it to the binding energy difference of $\rm ^{12}CO$ and $\rm ^{13}CO$ to ices.
However, it is unclear if the low \icn/\jcn\ ratio at $r\sim100$ au is actually correlated to the high $\rm ^{12}CO/^{13}CO$ ratio beyond $r\sim100$ au since the spatial resolution of the data in \citet{yosh22} and this paper are both limited.
The vertical variation of the \icn/\jcn\ ratio can be another explanation.
In terms of the $\rm H_2$ volume density, the CN emission at $r\sim100$ au arises from a deeper region compared to $30 < r < 80$.
Therefore, if the \icn/\jcn\ ratio near the midplane is lower than that in the {upper layer}, the observed profile could be reproduced without significant radial variation.
We note that these two possibilities are not mutually exclusive; both radial and vertical differences in the \icn/\jcn\ ratio may contribute to it.
In both cases, the physical structure such as the gap seen in the scattered light \citep{boek17} may be related to the variation of \icn/\jcn.

\subsection{ {  Caveats} }
{  
There are some caveats for our analysis.
First, our non-LTE modeling uses the RADEX code, which does not treat e.g. the overlap effect of emission lines. This might affect the excitation conditions as well as the resultant parameters.
However, this effect is beyond the current radiative transfer capabilities. }

We should also pay attention to the non-LTE spectra model fitting, where we assume that the kinetic temperature and $\rm H_2$ volume density are the same among all hyperfine lines.
This means, in other words, we assume that the emission comes from a homogeneous medium along a line-of-sight.
However, the temperature and density structures in protoplanetary disks have a vertical gradient in reality \citep[e.g.,][]{dull02, law21}.
Therefore, if the kinetic temperature of the line emitting regions of \icn\ and \jcn\ are different, it may be possible that the retrieved \jcn\ column density is affected by the optically thick \icn\ hyper-fine lines.
More detailed modeling is required to exclude this possibility.
However, it is not only beyond the scope of this paper but also it needs deeper observations with multiple transitions of \jcn.

\section{Summary} \label{sec:sum}

We have analyzed archival ALMA data of the \jcn\ lines in the TW Hya disk. Our findings are summarized as follows:

\begin{enumerate}
\item The \jcn\ $N=2-1$ hyperfine lines are detected.
This is the first spatially-resolved detection of \jcn\ in protoplanetary disks.
The ring-like morphology of the emission resembles that of \icn.

\item In conjunction with the \icn\ lines, we conducted a non-LTE analysis to derive the kinetic temperature and ${\rm H_2}$ volume density of the line emitting regions, and \icn\ and \jcn\ column densities at each disk radius.
The derived low ${\rm H_2}$ density of $\sim 4\times10^7\ {\rm cm^{-3}}$ at $r < 80$ au suggests that the CN traces the disk upper layer ($z/r\sim 0.2-0.3$).

\item The \icn/\jcn\ ratio is derived to be $70^{+9}_{-6}$ between $30 < r < 80$ au, which is similar to the ISM value but different from the previous measurements of carbon isotope ratios of CO and HCN in the TW Hya disk, motivating the following scenario.
First, element oxygen in gas-phase was depleted with respect to carbon, which promoted the isotope exchange reaction (\ref{eq:CO}). This produced $\rm ^{12}C^{+}$ which was eventually locked to dust grains, making the gas phase $\rm ^{13}C$-rich.
The relatively $\rm ^{13}C$-poor CN isotopologue ratio could be explained by the in-situ reactions including reactions (\ref{eq:CO})-(\ref{eq:CN3}).


\item \icn/\jcn\ tentatively shows a low value of $48^{+5}_{-4}$ at $r\sim 100$ au. This different in ratio can be explained due to the CN lines tracing different altitudes at different radii, where the isotope chemistry differs.

\item This study highlights that different carbon-bearing species have different carbon isotope ratios in protoplanetary disk gas. Such isotope fractionation patterns would affect the isotope composition of the planetary system material.

\end{enumerate}

\begin{acknowledgments}
We would like to acknowledge the anonymous referees for their helpful remarks and comments. 
We also thank to Dr. Kaori Kobayashi for helpful discussion on the spectroscopic data.
This paper makes use of the following ALMA data: ADS/JAO.ALMA\#2016.1.01375.S, 2018.A.00021.S, and 2017.1.01199.S.
ALMA is a partnership of ESO (representing its member states), NSF (USA), and NINS (Japan), together with NRC (Canada), MOST and ASIAA (Taiwan), and KASI (Republic of Korea), in cooperation with the Republic of Chile.
The Joint ALMA Observatory is operated by ESO, AUI/NRAO, and NAOJ.
This research has made use of spectroscopic and collisional data from the EMAA database (\url{https://emaa.osug.fr} and \url{https://dx.doi.org/10.17178/EMAA}). EMAA is supported by the Observatoire des Sciences de l’Univers de Grenoble (OSUG)
This work was supported by Grant-in-Aid for JSPS Fellows, JP23KJ1008 (T.C.Y.).
This work was partially supported by Overseas Travel Fund for Students (2022) of Department of Astronomical Science, The Graduate University for Advanced Studies, SOKENDAI.
T.C.Y. was also supported by the ALMA Japan Research Grant of NAOJ ALMA Project, NAOJ-ALMA-306.
H.N. is supported by JSPS KAKENHI grant numbers 18H05441, 19K03910, 20H00182.
K.F. is supported by JSPS KAKENHI grant numbers 20H05847 and 21K13967.
S.L. is supported by a Korea Astronomy and Space Science Institute grant funded by the Korean government (MSIT) (Project No. 2024-1-841-00).
Support for C.J.L. was provided by NASA through the NASA Hubble Fellowship grant No. HST-HF2-51535.001-A awarded by the Space Telescope Science Institute, which is operated by the Association of Universities for Research in Astronomy, Inc., for NASA, under contract NAS5-26555. CHR acknowledge the support of the Deutsche Forschungsgemeinschaft (DFG, German Research Foundation) Research Unit ``Transition discs'' - 325594231. CHR is grateful for support from the Max Planck Society.

\end{acknowledgments}

%

\vspace{5mm}
\facilities{ALMA}


\software{astropy \citep{astropy:2013, astropy:2018, astropy:2022},  CASA \citep{mcmu07}, emcee \citep{fore13}, Keplerian\_mask \citep{kepmask}, SpectralRadex \citep{hold21}, gofish \citep{teag16} }

\begin{figure*}[hbtp]
    \epsscale{1.1}
    \plotone{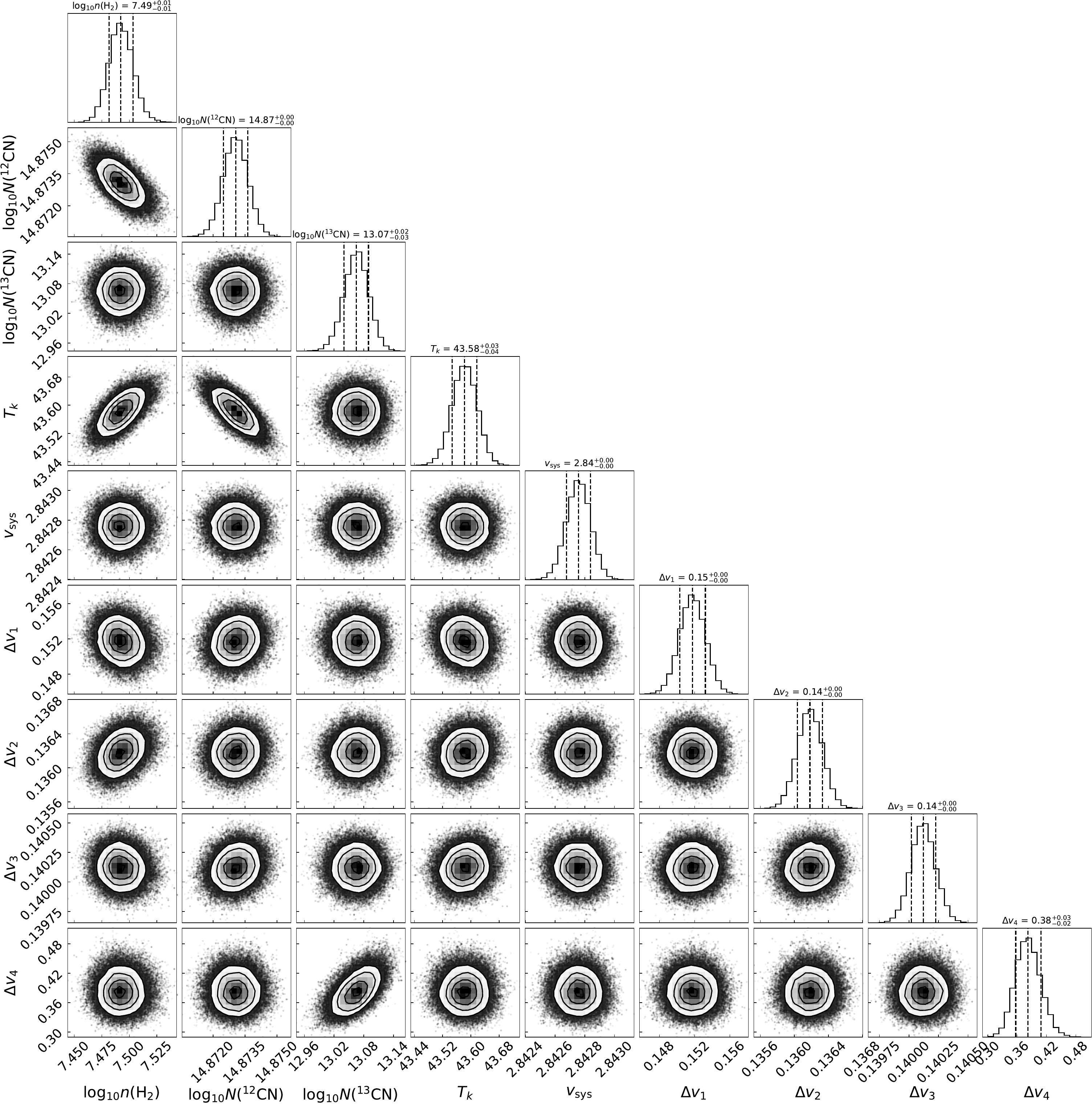}
    \caption{ Marginal posterior distribution of the fitting parameters at $r=56$ au. The units of each parameter are given in Table \ref{tab:prior}. The subscripts of the last four parameters indicate the SPWs for ${\rm ^{12}CN\ N=2-1,\ J=3/2-1/2}$, ${\rm ^{12}CN\ N=2-1,\ J=5/2-3/2}$, ${\rm ^{12}CN\ N=1-0}$, and ${\rm ^{13}CN\ N=2-1}$ lines, respectively.}
    \label{fig:c1}
\end{figure*}

\begin{figure*}[hbtp]
    \epsscale{1.1}
    \plotone{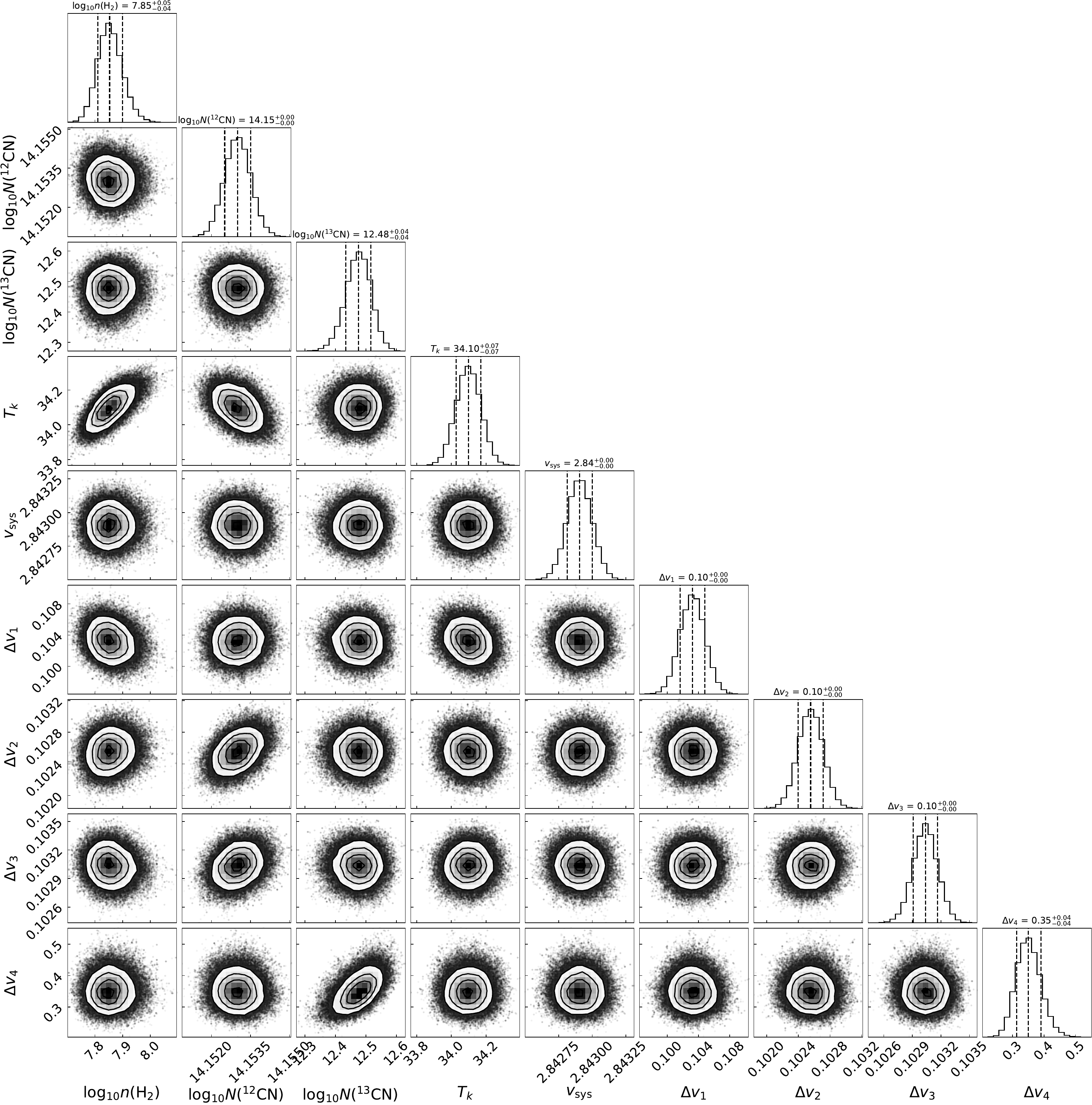}
    \caption{ Same as Figure \ref{fig:c1}, but for the results at $r=101$ au.  }
    \label{fig:c2}
\end{figure*}

\appendix
\label{appendix}
The corner plots of the resultant marginal posterior distribution at $r=56$ au and $101$ au are shown in Figure \ref{fig:c1} and \ref{fig:c2}, respectively.

\bibliography{sample631}{}
\bibliographystyle{aasjournal}



\end{document}